\documentclass[runningheads]{svmult}

\usepackage{makeidx}   
\usepackage{graphicx}  
\usepackage{subeqnar}  
\usepackage{multicol}  
\usepackage{physprbb}  
\makeindex             

\newcommand{\Msolarcjw}{\mbox{\,$\rm M_{\odot}$}}        
 
\newcommand{\gtsimeqcjw}{\raisebox{-0.6ex}{$\,\stackrel 
        {\raisebox{-.2ex}{$\textstyle >$}}{\sim}\,$}} 
\newcommand{\aseccjw}{^{\prime\prime}}

\begin{document}
\title*{A search for the first massive galaxy clusters}
\toctitle{A search for the first massive galaxy clusters}
%
%
\titlerunning{A search for the first massive galaxy clusters}
%
\author{Chris J.\ Willott\inst{1}
\and David Crampton\inst{1}
\and John B.\ Hutchings\inst{1}
\and Marcin Sawicki\inst{1}
\and Luc Simard\inst{1}
\and Matt J.\ Jarvis\inst{2}
\and Ross J.\ McLure\inst{3}
\and Will J.\ Percival\inst{3}}
\authorrunning{Chris J.\ Willott et al.}
%
%
\institute{Herzberg Institute of Astrophysics, National Research Council,
5071 West Saanich Rd, Victoria, B.C. V9E 2E7, Canada
\and Astrophysics, Department of Physics, Keble Road, Oxford, OX1
3RH, U.K.
\and Institute for Astronomy, University of Edinburgh, Royal
Observatory,  Blackford Hill, Edinburgh, EH9 3HJ, U.K.}

\maketitle              

\begin{abstract}
We have obtained deep, multi-band imaging observations around three of
the most distant known quasars at redshifts $z>6$. Standard accretion
theory predicts that the supermassive black holes present in these
quasars were formed at a very early epoch. If a correlation between
black hole mass and dark matter halo mass is present at these early
times, then these rare supermassive black holes will be located inside
the most massive dark matter halos. These are therefore ideal
locations to search for the first clusters of galaxies. We use the
Lyman-break technique to identify star-forming galaxies at high
redshifts. Our observations show no overdensity of star-forming
galaxies in the fields of these quasars.  The lack of (dust-free)
luminous starburst companions indicates that the quasars may be the
only massive galaxies in their vicinity undergoing a period of intense
activity.
\end{abstract}

\section{Introduction}
The existence of inactive supermassive black holes in the nuclei of
massive nearby galaxies (Magorrian et al. 1998) is a sure sign that
most galaxies underwent substantial active phases. Furthermore, the
correlation between the amount of material accreted (black hole mass)
and the galaxy mass indicates important links between galaxy formation
and AGN activity.  One implication is that the most massive black
holes live in the most massive galaxies which reside in the most
massive dark matter halos.

The $z>6$ quasars being discovered in the Sloan Digital Sky Survey
(see Fan, this volume) have exceptionally high luminosities
($M_{1450}<-27$). The available evidence suggests that these
luminosities are not amplified by gravitational lensing (Fan et
al. 2003) or beaming (Pentericci et al. 2002; Willott et
al. 2003). Assuming that these quasars are radiating at the Eddington
limit gives a lower limit on their black hole masses of several
billion solar masses. These are comparable to the black holes inside
the largest dominant cluster ellipticals at the present time.
Applying local calibrations from this black hole mass to galaxy and
halo masses suggests that these quasars reside in halos with mass
$\sim 10^{13} \Msolarcjw$ (e.g. Fan et al. 2001). LCDM simulations show
that structure grows `bottom-up' with the largest halos typically
collapsing at the latest epochs. Halos with mass $> 10^{13} \Msolarcjw$
at $z=6$ are therefore extremely rare (the fraction of mass in
such halos is of order $10^{-8}$; Sheth \& Tormen 1999)

There is evidence that at least some of these quasars are located in
massive galaxies. Sub-millimetre photometry of several of them show
very high far-infrared luminosities, implying extreme starbursts of $>
1000 \Msolarcjw\,{\rm yr}^{-1}$ (Bertoldi et al. 2003; Priddey et
al. 2003). The most distant quasar, SDSS\,J1148+5251 at $z=6.42$,
contains $ \gtsimeqcjw 10^{10} \Msolarcjw$ of molecular gas, again
indicative of a massive, primeval galaxy (Walter et al. 2003). These
distant quasars, which are thought to be residing in massive galaxies,
are therefore ideal places to search for the first clusters of
galaxies. The most straightforward method for identifying galaxies at
such high redshifts is the Lyman-break technique -- a large
discontinuity in the spectral slope due to absorption by neutral
hydrogen (e.g. Steidel et al. 1996). The increase in the optical depth
of hydrogen absorption at such high redshifts (e.g. Songaila \& Cowie
2002) makes the Lyman-break technique even more effective at $z\sim 6$
than at lower redshifts. To search for galaxies in these putative mass
overdensities, we carried out a search for Lyman-break
galaxies in fields of three $z>6.2$ SDSS quasars.

\section{Observations}

We have imaged three quasar fields with the GMOS-North imaging
spectrograph on the Gemini-North Telescope: SDSS\,J1030+0524 at
$z=6.28$, SDSS\,J1048+4637 at $z=6.23$ and SDSS\,J1148+5251 at
$z=6.42$. The imaging field-of-view is 5.5 arcmin on a side,
equivalent to a co-moving size of 13 Mpc at $z=6.3$. Observations
were carried out in queue mode during November and December 2003. The
typical seeing FWHM of the observations is in the range 0.5 to
$0.7\aseccjw$. Typical exposure times are $\approx 2$ hours in the
$z'$-band and $\approx 3$ hours in the $i'$-band. The relative
exposure times were designed to give similar sensitivity in the two
bands for very red objects with colours of $i'-z' \approx 1.5$.  Full
details of these observations and their reduction will be published
elsewhere.

\begin{figure}
\vspace{-0.3cm}
\begin{center}
\includegraphics[width=.9\textwidth]{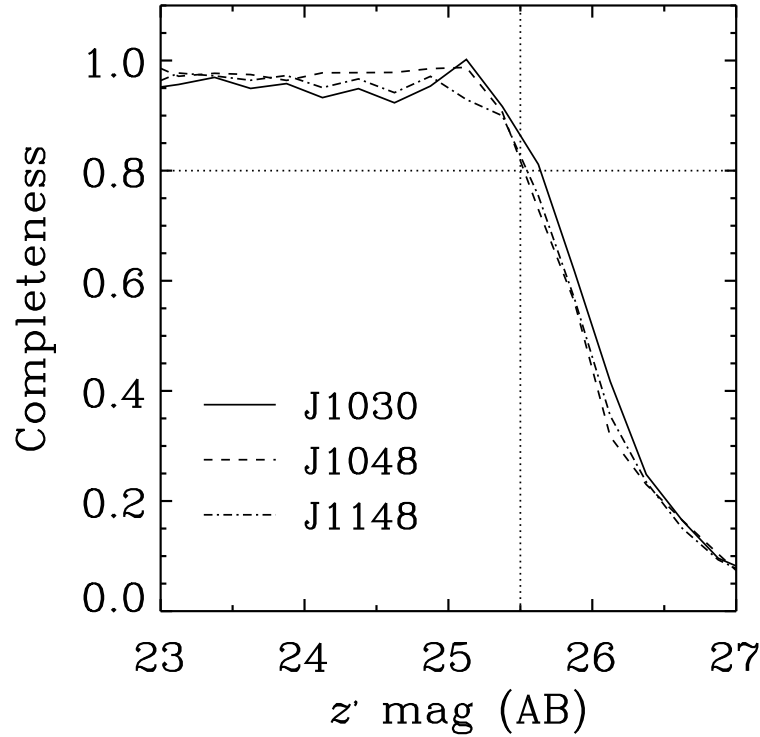}
\end{center}
\vspace{-0.9cm}
\caption[]{Completeness ratio against aperture-corrected $z'$-band
magnitude derived from recovery of simulated galaxies. The curves are
quite similar for the three different quasar fields. Dotted lines indicate
the location of a completeness ratio of 0.8 and the adopted complete
magnitude limit of $z'=25.5$.}
\label{fig:comp}
\end{figure}

Detection of objects in the reduced images was performed using the
Sextractor software (Bertin \& Arnouts 1996). The $z'$-band was
selected as the primary detection waveband since $z>6$ galaxies are
expected to have $i'-z'>2$ and may therefore be undetected at
$i'$. Sextractor was run in ``double-image'' mode to determine
$i'$-band measurements for objects detected in $z'$. Magnitudes on the
AB system were measured in circular apertures of diameter $1.5
\aseccjw$. Aperture corrections were applied statistically to
the $z'$-band magnitudes by fitting a linear function to the
difference between the total magnitude and aperture magnitude as a
function of aperture magnitude. Typical magnitude limits ($3\sigma$
limits in $1.5 \aseccjw$ apertures) are $z'\approx 26.2$ and $i'\approx
27.6$.

To assess the completeness of the $z'$-band catalogues we consider
both the observed number counts and the recovery of simulated objects.
The number counts in the three fields do not differ significantly from
each other. They agree well with the $z'$-band counts determined from
a much larger area survey (0.2 square degrees or thirty times the
GMOS-North field-of-view) by Capak et al. (2004). The number counts in
the quasar fields begin to change slope at $z'>25.5$ and turn over at
$z'=26$, indicating this is where the sample becomes incomplete.

The source recovery as a function of magnitude was determined by
populating the images with artificial galaxies and then using
Sextractor to attempt to detect these objects. About 10\,000
artificial galaxies with magnitudes in the range $23<z'<27$ were
placed into copies of the $z'$ images of each quasar. Regions of the
images already occupied by objects were masked out of the process to
eliminate incompleteness due to blending. The resulting completeness
ratio is plotted as a function of magnitude for the three quasar
fields in Fig.\,\ref{fig:comp}. The completeness in all the fields is
fairly flat at close to 1 up to $z'=25.2$ and then begins to
decline. The rapid decline occurs at $z'>25.5$ and the completeness
drops to 0.5 by $z=26.0$. All the fields have completeness $>0.8$ at
$z'=25.5$ and we adopt this as the magnitude at which completeness
begins to become an issue. This analysis with simulated objects agrees
well with the results for the number counts discussed previously.

\section{High-redshift galaxies in the quasar fields}
\label{search}

The $z>6$ SDSS quasars have $z'\approx 20$ and colours of
$i'-z'=3.25,2.98,3.25$ respectively for J1030, J1048 and J1148.
Quasars and star-forming galaxies have broadly similar spectra over
the rest-frame wavelength range $90-140$\,nm probed by the $i'$ and
$z'$ filters. Their spectra are dominated by a large break due to
absorption by neutral hydrogen. Therefore one would expect companion
galaxies to have comparable colours to the quasars. 

\begin{figure}
\vspace{-0.3cm}
\begin{center}
\includegraphics[width=.99\textwidth]{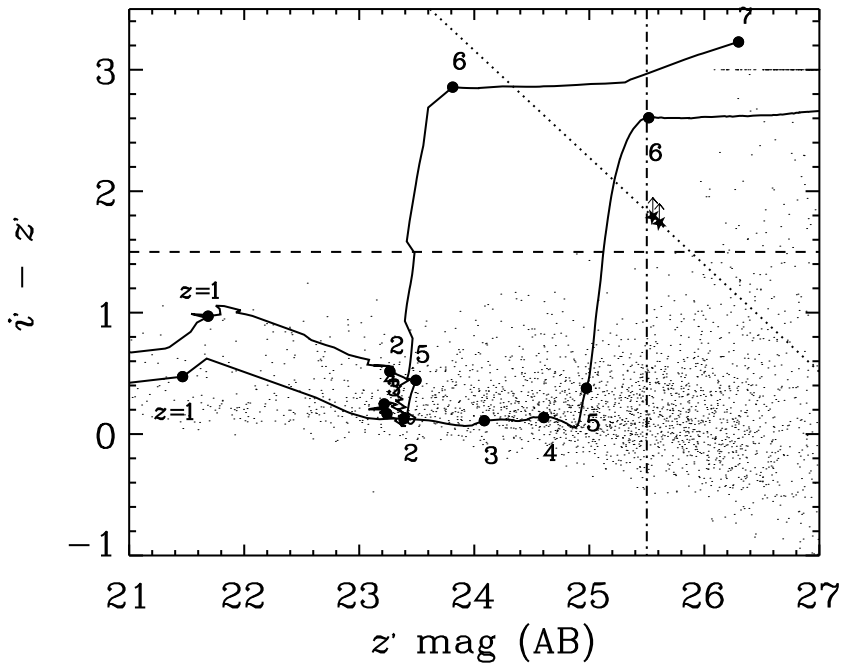}
\end{center}
\vspace{-0.3cm}
\caption[]{Colour-magnitude diagram for objects detected at $z'$-band
in the field of the $z=6.28$ quasar SDSS\,J1030+0524. The completeness
limit of $z'=25.5$ is shown with a dot-dashed line. The high-redshift
galaxy selection criterion of $i'-z'>1.5$ is marked with the dashed
line. The dotted line represents the colour of an object which is just
detected at the $3\sigma$ level in the $i'$-band as a function of the
$z'$-band magnitude.  Most objects with measured zero or negative flux
in the $i'$-band are plotted at $i'-z'=3$. The exceptions to this are
objects which pass the $i'$-band dropout selection criteria discussed
in the text.  These are shown with star symbols and lower limits on
the $i'$-band $3\sigma$ line. The dots at $i'-z'>1.5$ indicate sources
which are detected at less than $4\sigma$ at $z'$-band and hence have
very large uncertainties on their magnitudes and colours and in some
cases may be spurious.  The labelled curves show the colour and
magnitude as a function of redshift for an evolving L$^{\star}$
elliptical (upper curve) and a non-evolving L$^{\star}$ Lyman-break
galaxy (lower curve) -- see text for more details.}
\label{fig:colmags}
\end{figure}

A colour-magnitude diagram for objects in the field of
SDSS\,J1030+0524 is shown in Fig.\,\ref{fig:colmags}. Most objects
have colours in the range $0<i'-z'<1$ as is well known from previous
surveys (Dickinson et al. 2004; Capak et al. 2004). Also plotted are
curves showing the colour-magnitude relation for two different types
of galaxy. Model galaxy spectra were generated from the Bruzual \&
Charlot (2003) spectral synthesis code with Lyman forest absorption
evolution matching the observations of Songaila \& Cowie (2002). The
upper curve is a present-day L$^{\star}$ elliptical which formed all
of its stars in a starburst starting at $z=10$ with a characteristic
timescale of 1\,Gyr and evolved since without merging. Note
that there is no dust extinction assumed for this model, but in
reality the dust extinction would increase with redshift (due to
evolution and $k$-correction) making the galaxy fainter than plotted
at higher redshifts. The lower curve is a L$^{\star}$ Lyman-break
galaxy model where the galaxy is observed 0.5\,Gyr into a constant
star formation rate starburst.

A search was made for objects which could plausibly be high-redshift
galaxies. The $i'$-band dropout selection criteria adopted were snr in
the $z'$-band $\ge 4$ and a colour of $i'-z' \ge 1.5$. Possible
candidates were inspected and magnitudes checked to ensure their
unusual colours are not spurious. A total of four objects satisfying
these criteria were found; two in the field of SDSS\,J1030+0524, shown
in Fig.\,\ref{fig:colmags} with filled symbols, and one in each of the
other two fields. Only one of these four galaxies has a magnitude
brighter than the completeness limit of $z'=25.5$. This object has
$i'-z'=1.56 \pm 0.31$ which is quite close to the colour selection
value and the size of the uncertainty means it is quite plausible that
photometric errors have scattered the colour into the dropout
range. The measured colour is about $5\sigma$ away from the colours of
the SDSS quasars, which suggests that if it is a high-redshift galaxy,
then it is most likely foreground to the quasars with a redshift in
the range $5.7<z<6$. The other three objects have $z' \approx 25.6$
and only lower limits on their $i'-z'$ colours which lie in the range
$1.7-1.9$.

Our $z'$-band images are complete at the greater than 80\% level to
$z'=25.5$. At $z' \le 25.5$, every single $z'$-band object detected on
our images has a counterpart at the $>3\sigma$ level in the $i'$-band
image. At fainter magnitudes, this is no longer true and our
constraints on the $i'-z'$ colours of objects at these magnitudes
becomes very weak due to uncertainty in both the $z'$ and $i'$
magnitudes.  We now consider the number of $i'$-band dropouts we could
have expected to find under the assumption that the quasar fields are
``random'' and show no enhancement due to the existence of the
quasars. The best comparison dataset which goes deep enough over a
wide area is the {\it Hubble Space Telescope} ACS imaging of the GOODS
regions (Giavalisco et al. 2004). These observations give a surface
density of objects with $z'<25.5$ and $i'-z'>1.5$ of 0.02
arcmin$^{-2}$ (Dickinson et al 2004; Bouwens et al. 2004). The total
sky area we have surveyed with GMOS-North is 82 arcmin$^{2}$.
Therefore on the basis of the GOODS observations we would expect 1.6
$i'$-band dropouts in our total area. Our finding of one dropout is
entirely consistent with the expectations for blank fields.

Our observations show that these quasar fields do not exhibit an
excess of luminous companion galaxies. The magnitude limit of
$z'=25.5$ corresponds to a UV luminosity of
$L_{1500}=2.5\times10^{29}{\rm erg s}^{-1}{\rm Hz}^{-1}$ at a redshift
of $z=6.3$. This is equivalent to an unobscured star formation rate of
$SFR=30\, \Msolarcjw\, {\rm yr}^{-1}$, assuming the conversion given in
Madau, Pozzetti \& Dickinson (1998). The few known galaxies at
redshifts $z\approx 6.6$ discovered in narrow-band surveys have star
formation rates derived from their UV luminosities comparable to this
limit (Hu et al. 2002; Kodaira et al. 2003). For comparison, the
millimeter detections of dust in J1048 and J1148 imply total star
formation rates $> 1000\, \Msolarcjw {\rm yr}^{-1}$ in the host galaxies
of the quasars (Bertoldi et al. 2003).

How do we explain the lack of companion galaxies with high star
formation rates? One possibility is that galaxies undergoing intense
star formation do exist but they are heavily extinguished by
dust. Such galaxies may be detectable via their far-infrared emission
and we are conducting a program of sub-mm imaging around these quasars
to identify such galaxies. Deep observations with the IRAC camera on
the {\it Spitzer Space Telescope} would also be able to reveal these
obscured galaxies. Another possibility is that massive companion
galaxies exist but they are not observed during a period of intense
star formation. This possibility seems unlikely since the youth of the
universe at this redshift means galaxies would likely be gas-rich and
forming stars, especially if located in a dense environment which is
not yet virialized. Finally, we raise a question mark over the notion
that these rare massive black holes actually reside in the rarest,
densest peaks in the dark matter distribution. Perhaps the correlation
between black hole mass and dark matter halo mass (or halo circular
velocity -- see Wyithe \& Loeb 2004) does not apply to these quasars
and they are actually located inside less massive dark matter halos.

%

\end{document}